# In Situ Simultaneous Measuring Forces Acted on Individual Legs of Water Striders According to Archimedes' Principle

Yelong Zheng[‡], Hongyu Lu[‡], Wei Yin, Dashuai Tao, Lichun Shi, Yu Tian*

*State Key Laboratory of Tribology, Department of Mechanical Engineering, Tsinghua University, Beijing 100084, China*

‡ These authors contributed equally to this work.

\* Correspondence should be addressed to [tianyu@tsinghua.edu.cn](tianyu@tsinghua.edu.cn)

The supporting forces of all the legs have been simultaneously measured through a natural phenomenon inspired shadow method, by capturing water strider leg shadows to calculate the expelled water volume and the equivalent floating forces according to Archimedes' principle, with a sensitivity up to nN~pN/pixel. The method was verified by comparing to electronic balance to weight water striders. The supporting force acted on legs was found to linearly proportional to its shadow area, which geometry clearly indicated the hydrophobicity of legs. The pressed depth of legs, vertical weight focus position change and body pitch angle during leg refreshing, sculling and jumping forward of water striders have been achieved with resolutions of 5 μm/pixel, 2 μm/pixel, and 0.02°, respectively, which is difficult for general imaging technologies. The shadow method was also developed into a force measuring device and characterized the adhesion force of a hair/polymer surface contact in a few μN with a 0.3 nN/pixel resolution.

In situ simultaneous measuring forces acted on individual legs of animal and insects in natural state is difficult and crucial for disclosing their locomotion principles to guide the design of advanced biomimetic robotics. Water-walking arthropods, such as water strider, usually press water surface downward to support their weight in dynes[1–3], and scull to move on water surfaces[4–10]. However, the supporting forces of all the legs have not been simultaneously measured yet. Force measurement plays a crucial role in human manufacturing, trading and natural principle exploration. Besides general force measurement methods based on lever balance, strain deformation, objects motion etc., more than two thousand years ago, the ancient Greek scientist Archimedes found that "any floating object displaces its own weight of fluid", later called the Archimedes' principle[11]. In the Three Kingdoms period of China, a child prodigy, Chong Cao also proposed similar ingenious method of using the buoyancy principle to weigh an elephant[12]. Many modern water or air floating vehicles, such as ship, submarine, and dirigible, are designed according to the principle of flotation. Besides those giant vehicles, along with the development of micro/nano manufacturing, the size of actuators and sensors has been significantly reduced[13, 14]. Considering the exceptional capability of water-walking arthropods, there are lots of interest to explore their floating principles to establish biomimetic water-walking robots[7–18], including their hydrophobic waxy and hairy legs to reside on water surfaces[7, 19, 20] and the hydrodynamic propulsions by transfer momentum through capillary waves and hemispherical vortices by sculling their legs[4, 6, 8].

In former studies, the weight of a typical water-walking arthropods, water strider, was thought to be supported by the water pressure and the surface tension forces acted on legs[1, 4, 7, 15] as $Mg=F_b+F_c$, where $F_b$ is hydrostatic pressure acted on the leg area in contact with water, $F_c$ is the vertical component of the surface tension, $\sigma sin\theta$, along the contact perimeter[4]. Therefore, the pressed depth, contact area, contact profile, and contact angle of legs contacted with water should be simultaneously measured to achieve the supporting force. Weak forces in nN to μN in ambient environment are usually achieved with atomic force microscopy (AFM)[21–23], which could not be used for live water striders. Another way is to measuring forces acted on legs according to Archimedes' principle, the weight of the volume of water repelled by a floating object is equal to its own weight[5, 24, 25]. An ex vivo measurement with electronic balance showed that the maximum supporting force of one water strider leg could be up to 152 dynes (1 dyn = 10 μN), several times of its body weight[7]. A recent study showed some force estimations of the jumping upward of water striders and robotic insects[24]. However, a simultaneous image recording of all the legs from side view is difficult and generally low resolution due to the legs distributed at cm in standing width, mm in compressed depth. Therefore, up to now, a simple comparison between Archimedes' principle based force measurement to the static weight of a water strider has not been realized yet. An in vivo simultaneous force measuring of individual legs of water-walking arthropods, such as water strider is still a key challenge to be solved. In this study, we proposed a simple and high sensitive method that well overcomes the challenge.

Water striders usually left clear rounded or half-spindle shaped black shallows of their legs on the bottom of ponds with bright rings outside under sunshine as shown in Fig. 1a. The strange shapes and brightness of the shadow of their legs are due to the diffraction of the distorted water surfaces pressed by their legs. The size of leg shadows usually shows a positive relationship to the size/weight of water striders. Based on the Archimedes' principle, if the expelled water volume of legs of water strider could be measured, one can achieve the floating forces acted on individual legs and weigh a water strider. Therefore, we designed a test rig as shown in Fig. 1b, a point light irradiated the water strider residing above water surface to form shadows on the bottom of vessel, which was recorded with a camera. A typical top view of a water strider (out of focus, upper one) and its shadow (focused, lower one) is shown in Fig. 1c. When a hydrophobic slim pole, such as one leg of water strider leg, is slowly pressed down, the curvature of water surface is mainly governed by the gravity pressure depending on the water depth and the Laplace pressure determined by water surface curvature and surface tension[4–8, 26-27]. According to former researches[4–8], the curvature of water surface pressed by a slim pole at different depths were calculated and shown in Fig. 1d. As the increase of pressed depth, the water surface profile becomes sharper, refracting light at a larger angle and forming a larger shadow width. Right outside the black shadow, there is a bright ring, which is due to the refraction of light by the distorted water surfaces and agrees with the linear optical analysis as shown in Fig. 1e. Using a linear-optical analysis software ZEMAX, the width of leg shadow $S$ at different pressed depth $h_0$ was numerically calculated as shown in Fig. 1f.

Through the shadow width one can derive the water curvature profile at each section of the shadow and calculate the section area of the repelled water. According to Archimedes' principle, the floating force strength corresponding to each shadow width could be calculated and also shown in Fig. 1f. For each leg shadow, by integrating all the shadow slices, the floating force acted on a leg could be achieved. The floating force strength approximately linearly increased with the increasing of $h_0$, qualitatively in agreement with the previous experimental and theoretical results[5, 25]. Then, the weight of a water strider is equal to the sum of forces acted on individual legs. With this method, the force measuring of individual legs of water striders was transferred to a plane shadow image recording, which could be simultaneously and conveniently accomplished.

To verify the effectiveness of the proposed shadow method, the weight of seven different water striders has been achieved using this method and compared with the results of an accurate electronic balance, as shown in Fig. 2a~c. The typical floating force, shadow area pixel, and the maximum pressed depth of individual legs of water strider No. V were shown in Fig. 2a. The force resolution (ratio of force to pixels) was approximately 10 nN/pixel. If the shadow of one leg corresponding to a floating force of 50 μN occupies one third area of 30 million pixels (such as Nikon D 810 camera), the force measurement sensitivity could be up to 5 pN/pixel, approaching the level of single molecular forces[21, 22]. The results in Fig. 2a also clearly showed that the ratios of floating forces to the maximum pressed depth for fore, middle and hind legs were different. Therefore, the maximum pressed depth of legs should not be simply linked to the floating force. A continuous picturing shadow of seven water striders in stilled state showed very stable values with a standard deviation of the weight measurement within 0.5 % as shown in Fig. 2b. The obtained weight measured with shadows agreed well with that tested with an electronic balance, which has a 0.1 mg precision and a 0.01 mg resolution, as shown in Fig. 2c. The deviations were less than 4 %, except for water strider No. VII was about 5.9 %. On the other hand, the weight focuses of the water striders have been calculated and indicated by red dots in Fig. 2a. They were right on the body center of water striders, showing the force and torque balances were well satisfied as expected.

Sometimes, on the condition that the leg shadows of water strider were not smooth, water striders would lift up its two fore-legs and rub them around their mouth, then rub one of them with their middle-legs or hind-legs as shown in Fig. 2d. During rubbing, the weight focus (red dots) was shifted to the right side. After rubbing the legs, the leg shadows became smoother, indicating that the hydrophobic state of the legs had been recovered. Considering the rubbing actions, the hydrophobic materials on legs should be came from secretions from their mouth and transferred to the legs, which could be called a leg refreshing process and has not widely noticed or discussed yet. The tiny forces acted on legs S1-S5 during the leg refreshing were shown in Fig. 2e. For the subtle action of leg S5, the force fluctuation was about 0.37-7.40 μN, which was mainly compensated by forces acted on leg S1 of about 100.57-94.94 μN. The whole supporting force acted on the water strider fluctuated slightly between 224.8 to 228.9 μN as shown in Fig. 2f. Assuming a rigid supporting by legs S2 and S3 due to their relatively stable supporting force values, the change of vertical position of body weight focus could be calculated from the maximum pressed depth of leg S1 (See methods). It changed in the range of -5 to 10 μm as shown in Fig. 2f, which could be very difficult to be achieved by using traditional displacement measuring methods on water striders.

The shadow method could also be used to quantitatively characterize and analyze various motions of water striders, such as sculling forward and jumping forward, as shown in Fig. 3. The shadow images, supporting forces $F_n$ acted on individual legs, the whole supporting forces, and the whole lateral propulsive force calculated[24] based on the hydrodynamic force and the surface tension force from the anisotropic shadow (see Methods) have also been achieved as shown in Fig. 3 b&d. During sculling forward of a water strider (33.1mg), assuming a rigid connection of its fore and hind legs with body, the shadows of legs S1, S2, S4 and S5, and the corresponding pitch angle (See methods) of its body of about ± 3° as shown in Fig. 3b. During jumping forward of an infant water strider (5.3 mg) as shown in Fig. 3c, the fore legs played an important role of cushioning in the process of landing on water surface. The sum of supporting force, the lateral force, and the pitch angle change of the body were shown in Fig. 3 b&d. The vertical and lateral forces and pitch angel change in jumping forward were more abrupt than that in sculling forward agree with our intuition. The disclosed motion principles are of significant importance to guide the design and control of sculling and jumping motions of advanced water-walking robots.

The shadow method could be also developed into a universal weak force measurement method. As an example, a surface force measurement system was established with a hydrophobic circular polytetrafluoroethylene (PTFE) circular plate (a thin polydimethylsiloxane (PDMS) surface on the top) as shown in Fig. 4. A simple calibration by putting known weights on the top of the PDMS surface has been done as shown in Fig. 4b. The fitted equation showed a force sensitivity of about 0.3 nN/pixel. To simulate a hair/human skin contact[28], the contact aging effect of a human hair with a small piece of smooth PDMS surface was investigated. The adhesion force increased significantly as the contact time increased from about 6 to 10 μN as shown in Fig. 4c, similar to behaviors of static friction forces in amorphous $SiO_2$ contacts[29].

As a summary, through the shadows of the hydrophobic objects, the distorted curvatures of water surface could be reconstructed to derive the expelled water volume, which represented the equivalent floating force. The force measurement sensitivity down to pN/pixel could be easily realized with general family using cameras with a time resolution depending on the frame rate of cameras. With the methods, some subtle behaviors of water striders have been clearly observed and quantitatively disclosed for the first time, which is very important to understanding their locomotion principles. It can be easily developed into surface force measuring apparatuses with high force sensitivities. This research provides a novel way to developing sensitive force measuring methods and could inspire other measuring principles and methods based on our observation in general lives.

**Methods**

**Preparation of water strider and shadow image recording.** Water striders were carefully captured with small fishing nets from the pond of Approaching Spring Garden at Tsinghua University, Beijing, China. They were raised in the acrylic aquarium in the laboratory at room temperature (25 ℃), and were taken for experiment when needed. In shadow recording experiment, a water strider was put into an acrylic (PMMA: polymethyl methacrylate) vessel (10×15×20 cm$^3$) with a wall thickness of 5 mm, a water depth of 50 mm, and a white paper at the bottom of the vessel acting as a screen. A film ruler (transparent plastic ruler with thickness of 0.1 mm) was also placed at the surface of the screen. A 3 W white point light source (ZHPL-0803, Beijing Hezhong HangXun Sci. Tech. Corp. China) was placed 1000 mm right above the water surface. A Canon camera (EOS650D, maximum size: 5184 × 3456 pixels) was placed about 150 mm below the screen. Pictures or videos were recorded when the needed shadow of water striders came into central area of screen. The video mode is full HD recording with 1920 × 1080 pixels and 25 fps. A high-speed video camera (MQQ13MG-ON, Beijing Hezhong HangXun Sci. Tech. Corp. China) was also used to record videos with a size of 640 × 512 pixels during sculling forward and 512 × 404 pixels for jumping forward. The camera was placed above the water surface in recording jumping forward. The pictures or videos were read in Matlab program and processed in single pictures in shadow analysis.

**Calculation of distorted water surface curvatures.** According to former researches, the relationship between distorted curvatures of water surface pressed by a leg of water strider at a certain point and different pressed depths of the leg could be given by the following equations[30],

$$h(x, \emptyset, \theta) = B(\emptyset, \theta_c) \cdot e^{-\frac{1}{c}x} \qquad (1)$$

$$B(\emptyset, \theta_c) = -c \cdot e^{(\frac{1}{c}R \sin \emptyset)} \cdot \tan(\theta_c + \emptyset - \pi) \qquad (2)$$

$$c = \sqrt{\frac{\sigma}{\rho g} \frac{1}{(1+s^2)^{\frac{3}{2}}}} \qquad (3)$$

where $\theta$ is the contact angle between leg and water, $\Phi$ is the submerge angle, σ is the surface tension of water (0.072 N/m), $B$ is area representing the surface tension force, $R$ is radius of a cylinder, $\rho$ is the density of water, g is the acceleration of gravity, $x$ is position on $x$-axis, $s$ is constant. The calculated curvatures at different depth were shown in Fig. 1d.

**Leg shadow area selection and tilt angle correction.** A typical leg shadow as shown in Fig. 1e has a bright region outside the dark shadow. In the shadow selection, actions with mouse to continuously click around the boundary of leg shadow were needed to help the area selection program and separate the shadow of the other body parts coded by Matlab script. A critical threshold of the gray was used to define the boundary of the dark leg shadow area, and separated from the background of the image. The point which brightness is lower than the threshold will be set to "1", otherwise to"0". Then the image was transformed into binary image for later process.

Actions with mouse to click the top and bottom points of leg shadows (shadow edges at both ends of the legs) was also carried out to help the program to calculate the tilt angle of shadow (between *y*-axis and top or bottom of the shadow). The leg shadow was tilted into 0° through rotation transformation by using the function (B = imrotate (A, angle, method) in the Matlab with the calculated tilt angle calculated to simplify the later floating force calculation of each leg.

**Vertical supporting force calculation according to Archimedes' Principle.** The supporting force strength could be calculated according to the width of the shadows. Each shadow width corresponds to a distorted water surface curvature. The area enclosed by the distorted water surface curvature and the above horizontal line of the free water surface presented the expelled water volume by the leg at this point. Integrating the shadow slices along the leg direction, the supporting force of a whole leg could be calculated. The images were extracted from the video by recording with the high-speed video camera. The time was decided by the order of the image. The leg shadows need to be selected with the help of researcher as described above. The dark leg shadow was divided into horizontal lines of pixels shadow, and integrated along *y* direction to calculate the floating force based on the obtained relationship between floating force strength to leg shadow width. The supporting forces acted on legs of water striders during static, leg refreshing, sculling forward and jumping forward have been calculated. The effect of leg radius on the shadow width could be neglected in this study. The small Weber number $W_e = \rho U^2 D/\sigma \sim 10^{-2}$ indicated that the dynamic pressure effect of the water curvature has been neglected when water striders were in various motions[24]. Here, $\rho$ was the density of water, $U \sim 0.1$ ms$^{-1}$ was the speed of leg, $D \sim 0.32$ mm was the leg diameter.

**The lateral force calculation.** $F_{s\_L}$ was lateral force originated from surface tension of the distorted water surfaces acted on the leg of water strider. The lateral surface tension strength, $f_{s\_L}$ was calculated by $f_{s\_L}= \sigma\cos(\alpha_a) - \sigma\cos(\alpha_p)$, where $\sigma$ was the surface tension of water, $\sigma=0.072$ N/m, $\alpha_a$ and $\alpha_p$ were anterior and posterior angles and calculated by $\alpha_a= \tan^{-1}(h_0/L_a)$ and $\alpha_p = \tan^{-1}(h_0/L_p)$, where $L_p$ and $L_a$ were the width of dimple slice, $h_0$ was depth of dimple slice [10]. To calculate the $\alpha_p$ and $\alpha_a$ through the distorted shadow conveniently, the light source was placed at 1 m above the water surface. The light source could be taken as the parallel light. The width of dimple slice was equal to the width of shadow. According to $L_a$ and Fig. 1f, $h_0$ was obtained. The lateral surface tension force, $F_{s\_L}$, was calculated by integrating $f_{s\_L}$ along y direction as

$$F_{s\_L} = \int_0^{L_{s\_L}} f_{s\_L}(y)dy \quad (4).$$

The lateral force caused by drag, $F_{d\_L}$ was obtained by $F_{d\_L}= 0.25\rho S_f U^2(1+10 R_e^{-2/3})$, where $S_f = \pi h_0 l/2$ was the frontal surface area of half a circular cylinder, $R_e = L_c U/\nu$ was Reynolds number, $\rho = 1$ g/cm$^3$ was the density of water, $h_0$ was depth of a dimple slice, $l$ was the length of leg, $\nu$ was the kinematic viscosity of the water, $\nu=1.01\times10^{-6}$ m$^2$/s, $L_c$ was the largest dimension parallel to the water flow and was estimated from shadow width of leg, $U$ was velocity of leg and obtained by the coordinate of fore leg and timer in camera[10].

**The vertical position change of weight focus and pitch angle of body calculation**

The vertical position change of weight focus of water striders, $h_{\text{weight foucs}}$ was estimated by $h_{\text{weight foucs}} = [d_1\times(h_{\text{maxs2}}+h_{\text{maxs3}})/2+h_{\text{maxs1}}\times d_2]/(d_1+d_2)$, where $h_{\text{maxs1}}$, $h_{\text{maxs2}}$ and $h_{\text{maxs3}}$ were maximum pressed depth of legs S1, S2 and S3 and calculated with the Fig. 1f and maximum width of the shadow, $d_1$ and $d_2$ were the distances from weight focus to the center of leg S1 and the center of connection between legs S2 and S3 respectively, $d_1 =11.2$ mm and $d_2 =19.5$ mm.

The pitch angle of body, $\varphi$ was calculated as $\varphi = \sin^{-1}[(h_{\text{maxs4}}+ h_{\text{maxs5}})/2L - (h_{\text{maxs1}}+ h_{\text{maxs2}})/2L]$, where $h_{\text{maxs1}}$, $h_{\text{maxs2}}$, $h_{\text{maxs4}}$, $h_{\text{maxs5}}$ were maximum pressed depth of legs S1, S2, S4 and S5, $L$ was length from the center of fore legs to the center of hind legs, $L = 18.14$ mm in sculling forward of water strider (33.1 mg), 12.56 mm for jumping forward of water strider (5.3 mg).

**Adhesion force measurement with a PTFE circular plate.** A circular PTFE sheet (diameter 12 mm, thickness 0.2 mm) stacked with a PDMS disk (diameter 5 mm, thickness 3 mm) on the top surface was used to float on water surface in the adhesion test. This adhesion force measurement system was calibrated by putting different weights on the top of the PDMS surface and recording the shadow area. Then it was used to test the adhesion strength between a hair and the PDMS surface. The one tested as shown in Fig. 4 was donated by a group member with a diameter 60 μm and a length 50 mm. It was glued on a one-dimensional translation stage drove by a stepper motor controller and a timer. At beginning, the test end of the hair was put about 0.5 mm above the PDMS surface. The hair was pressed down to contact with PDMS disc and separated at a driving speed of 0.05 mm/s. After the contact, the hair was lifted up after delaying time 30 s, 60 s and 90 s. The shadow of the lower disk was recorded by the video. The force curves were calculated by the above methods.

**Acknowledgements**

This work is supported by the National Natural Science Foundation of China (Grant Nos. 51425502 and 51323006).

# Figures

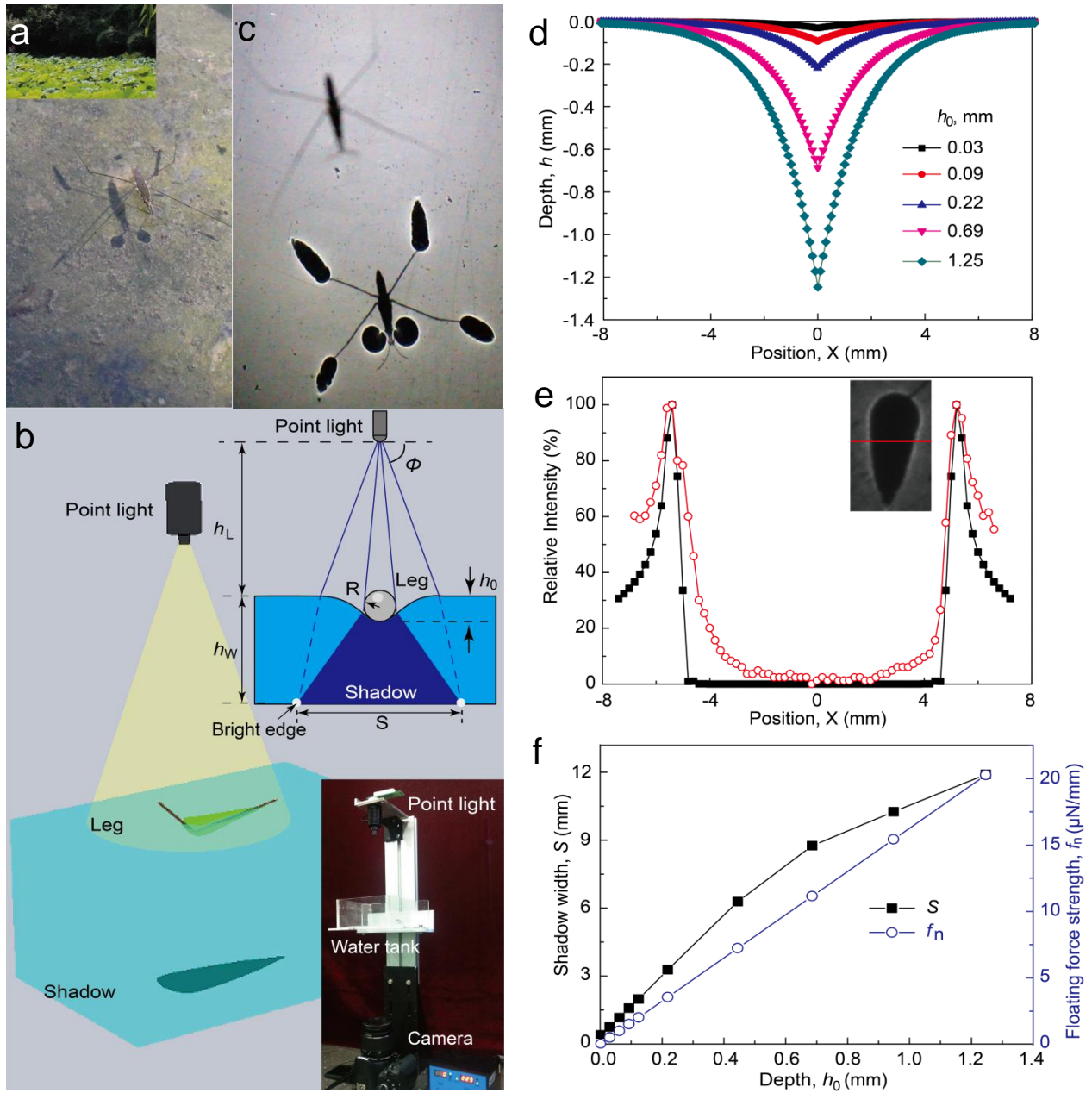

**Figure 1.** Method of measuring forces acted on legs of water strider from leg shadows. (a) A water strider and its shadow under sunshine in the pond of Approaching Spring Garden at Tsinghua University, Beijing, China. (b) The structure of the shadow recording apparatus. (c) A typical top view of a water strider shadow taken in laboratory. (d) Water surface profile section pressed by a hydrophobic slim pole at different depth. (e) Relative light intensity of a leg shadow section simulated with software and experimental results. (f) Shadow width and floating force strength of slim leg section pressed at different depth.

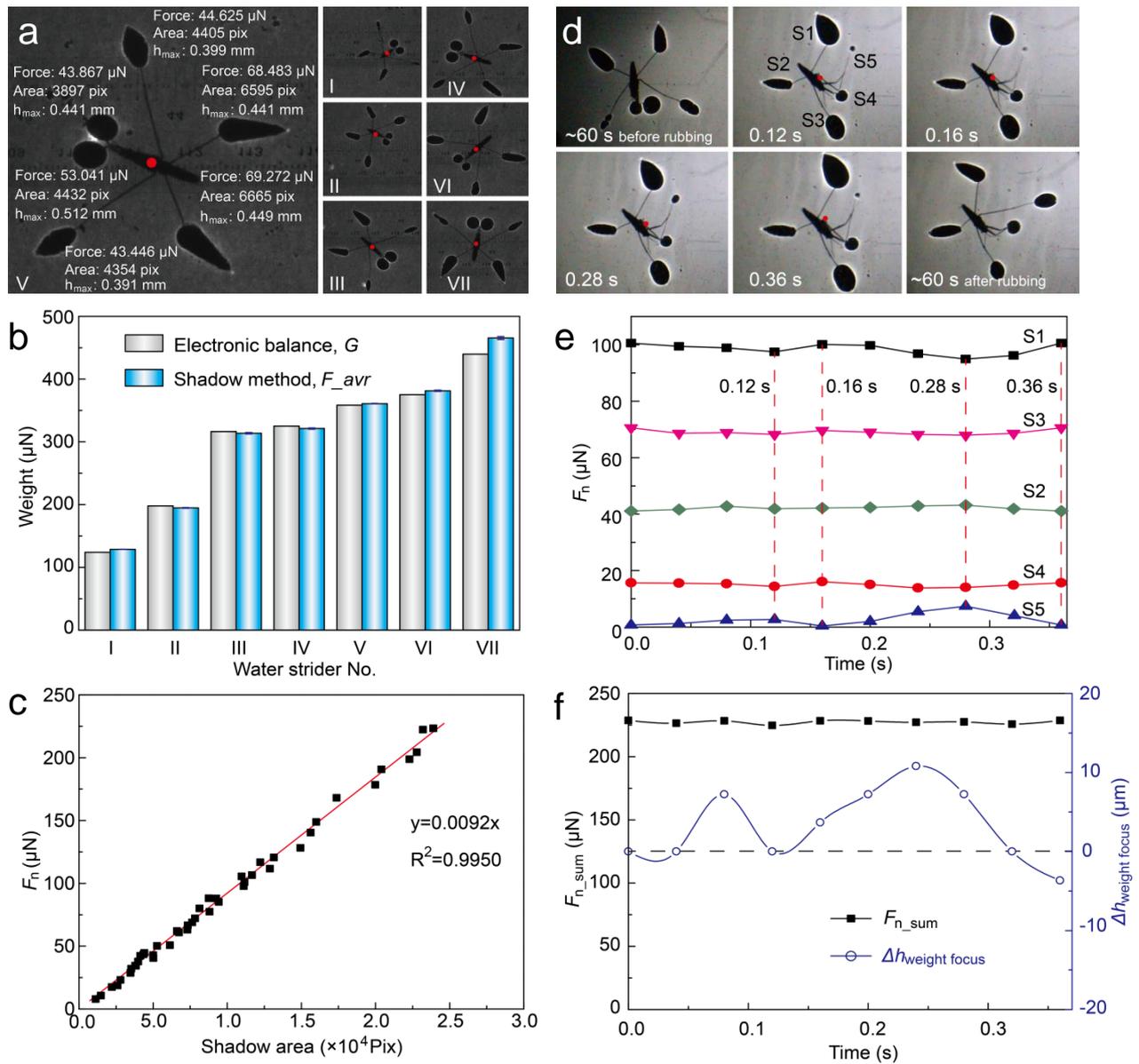

**Figure 2.** Simultaneous measuring supporting forces acted on legs of water striders in rest and leg refreshing. (a) Shadows images of seven water striders. (b) Comparison of weight measured from shadow experiment method and from electronic balance. (c) Relationship between the supporting force and the leg shadow area. (d) Shadow images of a water strider in its leg refreshing. (e) Forces acted on individual legs in leg refreshing process. (f) Sum of supporting forces of the water strider and the change of vertical position of its weight focus.

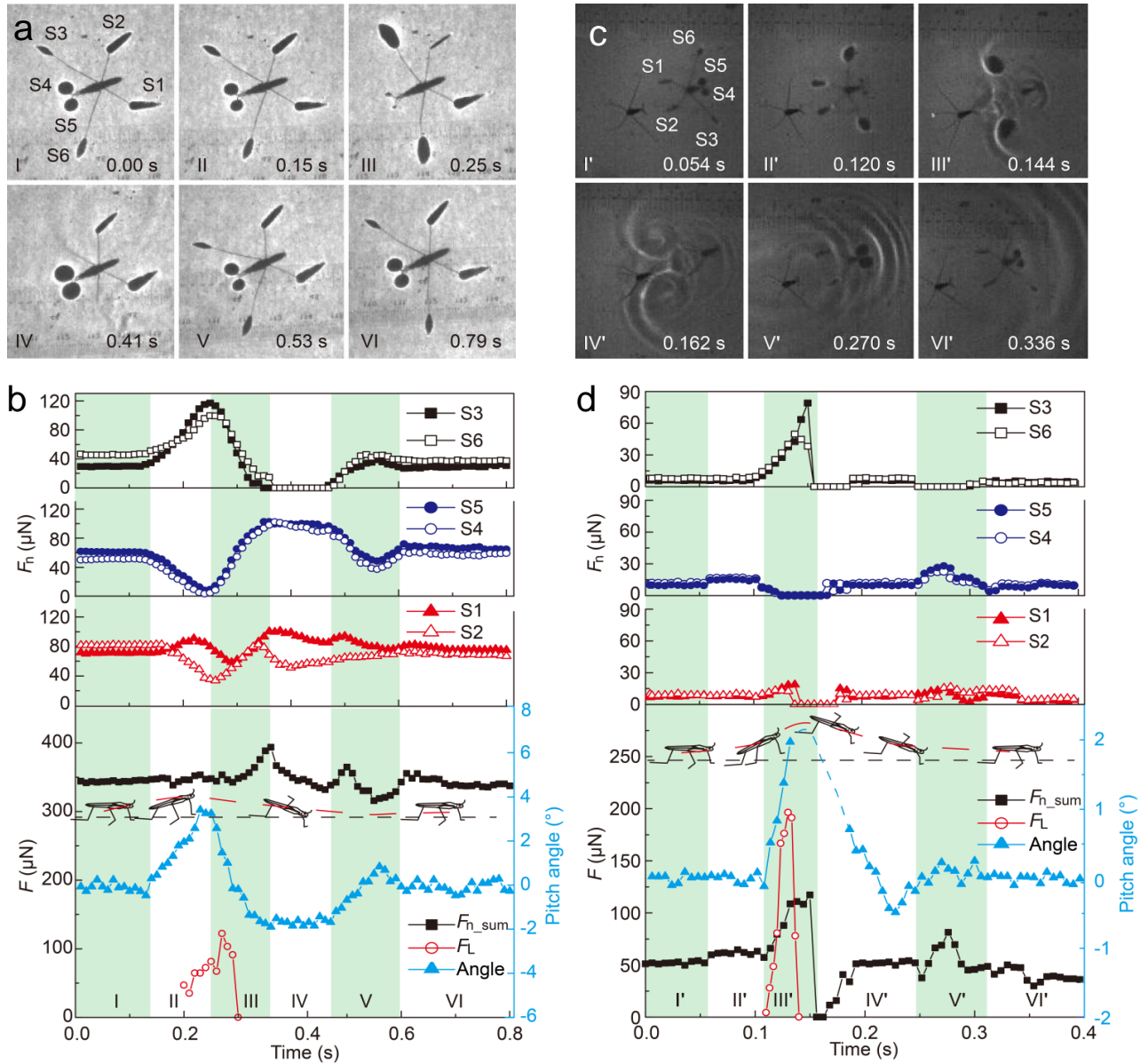

**Figure 3.** Forces acted on legs (numbered S1 to S6) of water striders in sculling forward (a, b) and jumping forward (c, d). Scale bars-rulers were on the background of the images. (a) Images of sculling forward. (b) Forces acted on individual legs, the sum of the vertical and lateral forces, and the pitch angle change of the body of water strider in sculling forward. **Sculling forward**: I-static state; II-start sculling: S3 and S6 were pressed down; III-sculling: S3 and S6 were pressed down hardly and sculled backward, and S4 and S5 were lifted slightly; IV-Post sculling: S3 and S6 were above water surface, S1, S2, S4 and S5 mainly supported the weight of its body; V- After sculling: S3 and S6 lied down on water surface again, six legs supported its body weight; VI-Rest state. (c) Images of jumping forward. (d) Forces acted on individual legs, the sum of the vertical and lateral forces, and the pitch angle change of the body of water strider in jumping forward. **Jumping forward**: I-static state; II- pre-jumping: S1, S2, S3 and S6 were pressed hardly on water surface; III- in jumping: S3 and S6 were pressed hardly on water surface, other legs were all lifted above water surface; IV-jumped up: all the legs were above water surface; V-landing: S4 and S5 hardly pushed water surface, S1 and S2 slightly landed on water surface, S3 and S6 were above water surface; VI-Rest state.

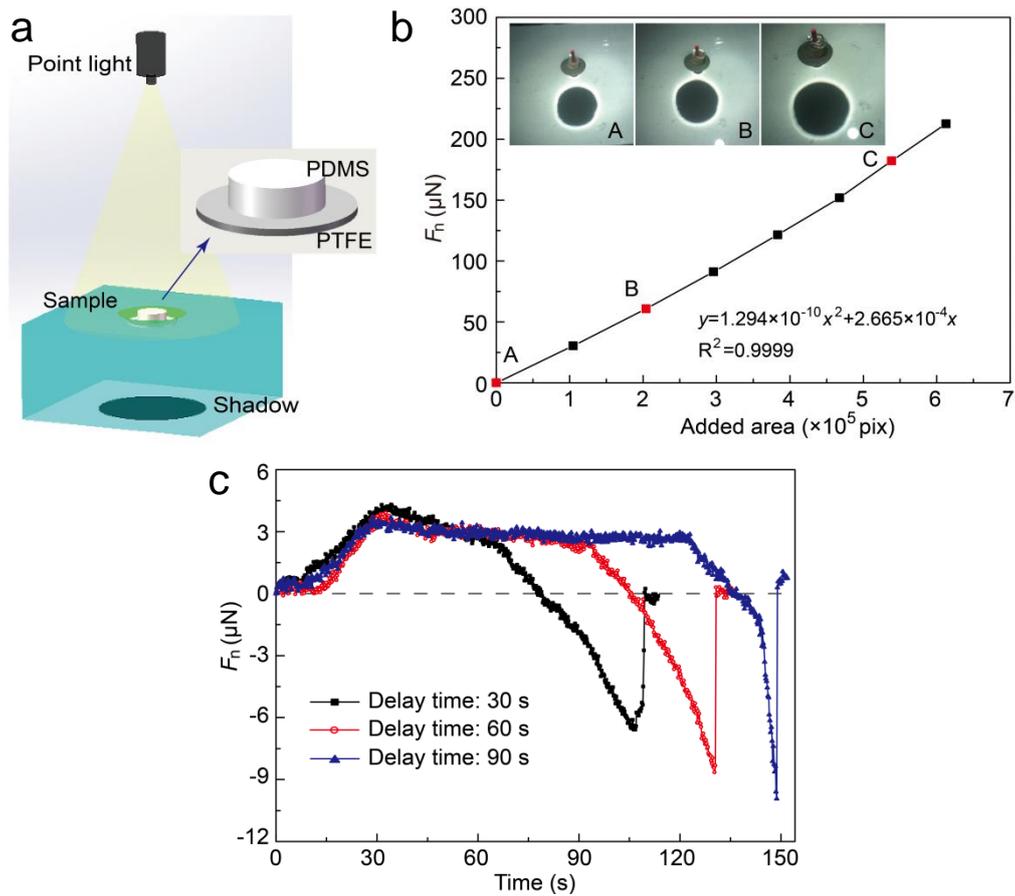

**Figure 4.** A simple apparatus based on shadow method for tiny adhesion force measurement. (a) Sketch of a PTFE circular plate based force measuring system. A point light source was shinned above, and a camera was used to record the shadow image on the bottom screen of water vessel. (b) Calibration of force and the shadow area. (c) Typical force-time curves of the adhesion test processes of a hair and a PDMS surface contact with a load force of about 4 μN applied for different period of time. The adhesion force increased significantly as the contact time increased.